\documentclass{aa}
\usepackage{graphicx}
\usepackage{txfonts}
\usepackage{tabularx}
\usepackage{multirow}
\begin{document}

\title{Vacuum-UV spectroscopy of interstellar ice analogs.} 
\subtitle{II. Absorption cross sections of nonpolar ice molecules.}
\author{G. A. Cruz-Diaz \inst{1}, G. M. Mu\~{n}oz Caro \inst{1}, Y.-J. Chen \inst{2,} \inst{3}, and T.-S. Yih \inst{3}}
\offprints{Gustavo A. Cruz-Diaz}
\institute{Centro de Astrobiolog\'{\i}a, INTA-CSIC, Carretera de Ajalvir, 
km 4, Torrej\'on de Ardoz, 28850 Madrid, Spain\\
email: cruzdga@cab.inta-csic.es, munozcg@cab.inta-csic.es 
\and Space Sciences Center and Depeartment of Physics and Astronomy, University of Southern California, Los Angeles, CA 90089-1341, USA
\and Department of Physics, National Central University, Jhongli City, Taoyuan Country 32054, Taiwan}
\date{Received - , 0000; Accepted - , 0000}
\authorrunning{G. A. Cruz-Diaz et al.}  
\titlerunning{Vacuum-UV spectroscopy of interstellar ice analogs}

\abstract
%Context
{Dust grains in cold circumstellar regions and dark-cloud interiors at 10-20 K are covered by ice mantles. A nonthermal desorption mechanism is invoked to explain the presence 
of gas-phase molecules in these environments, such as the photodesorption induced by irradiation of ice due to secondary ultraviolet photons. To quantify the effects of ice 
photoprocessing, an estimate of the photon absorption in ice mantles is required. In a recent work, we reported the vacuum-ultraviolet (VUV) absorption cross sections of nonpolar 
molecules in the solid phase.}
%Aims
{The aim was to estimate the VUV-absorption cross sections of nonpolar molecular ice components, including CH$_{4}$, CO$_{2}$, N$_{2}$, and O$_{2}$.}
%Methods
{The column densities of the ice samples deposited at 8 K were measured \emph{in situ} by infrared spectroscopy in transmittance. VUV spectra of the ice samples were collected 
in the 120-160 nm (10.33-7.74 eV) range using a commercial microwave-discharged hydrogen flow lamp.}
%Results
{We found that, as expected, solid N$_2$ has the lowest VUV-absorption cross section, which about three orders of magnitude lower than that of other species such as O$_2$, which is also 
homonuclear. Methane (CH$_4$) ice presents a high absorption near Ly-$\alpha$ (121.6 nm) and does not absorb below 148 nm. Estimating the ice absorption cross sections 
is essential for models of ice photoprocessing and allows estimating the ice photodesorption rates as the number of photodesorbed molecules per absorbed photon in the ice.}
%Conclusions
{}

\maketitle

\section{Introduction}

Grain surface chemistry is governed by the local H/H$_{2}$ ratio. If this ratio is high, hydrogenation is the main process, H$_2$O is the dominant constituent, and species
such as NH$_{3}$ and CH$_{4}$ are expected to form. On the other hand, if the H/H$_{2}$ ratio is substantially lower than unity, CO molecules will be abundant and species such as O$_{2}$ 
and N$_{2}$ are easily formed. Therefore, two types of ice mantles can be distinguished, one dominated by polar molecules and another one dominated by nonpolar, or only slightly 
polar, molecules (see Tielens et al. 1991; Chiar et al. 1995; Gerakines et al. 1996; Whittet et al. 1996). The relative to water abundances of CO$_2$ and CH$_{4}$ are between 
4-44\% and 0.4-8\%, respectively, for the different interstellar environments (Mumma \& Charnley 2011 and references therein).

Gas-phase molecular oxygen and nitrogen lack an electric dipole moment, which makes them infrared inactive. These species were therefore not detected in ice mantles and cannot 
be easily observed in the gas phase. In some clouds the estimated abundance of gaseous N$_2$ is on the order of 10$^{-5}$ with respect to H$_2$ (Bergin et al. 2002), whereas 
the coldest and densest cores show a decrease in the gas-phase N$_2$ by at least a factor of two (Belloche \& Andr\'e 2003). O$_2$ has a $^3 \Sigma$ ground state and thus a magnetic 
dipole moment enabling weak transitions in the submillimeter range between finestructure levels. Very low detection rates of O$_2$ were found toward Orion (Goldsmith et al. 2011) 
and the $\rho$-Oph A core (Liseau et al. 2012) with HIFI. The upper limits on O$_2$ and estimates of N$_2$ suggest relative abundances much lower than expected from the solar oxygen 
and nitrogen abundances. Large amounts of oxygen and nitrogen are therefore apparently missing from the gas phase and might be depleted on grains most likely in the form of O$_2$ 
and N$_2$ (Ehrenfreund \& van Dishoeck 1998).

Synchrotron radiation has been used as a source of VUV photons to perform VUV-absorption spectroscopy of ice samples. The National Synchrotron Radiation 
Research Center (NSRRC) in Taiwan (Lu et al. 2005, Cheng et al. 2011, and Wu et al. 2012), ASTRID at the University of Aarhus, 
and the UK Daresbury Synchrotron Radiation Source (Mason et al. 2006) were used as VUV sources in their measurements. This requires 
application for synchrotron beamtime, leading to a limited use, and one needs to work with a transportable chamber. The use of a microwave-discharge 
H$_{2}$ flow UV-lamp as the source for VUV-absorption spectroscopy, employed in various astrochemistry laboratories, allows a routine 
performance of VUV-absorption spectroscopy in the 120-170 nm spectral range. 
The polar molecular components of interstellar ice were studied in Cruz-Diaz et al. (2013a, Paper I). To present a more complete view of the 
interactions between the interstellar VUV field and icy grain mantles, we report here a similar study on nonpolar ices. Section ~\ref{Expe} 
describes the experimental protocol used in the experiments. Section ~\ref{VUV} summarizes the VUV-absorption cross-section measurements 
of the different nonpolar ices studied: CO$_{2}$, CH$_{4}$, N$_{2}$, and O$_{2}$. Gas-phase VUV-absorption cross-section data adapted from 
other works were used for comparison with our solid-phase data. In addition, the spectra of the different species were fitted with Gaussian 
profiles, using an in-house IDL code, to provide an individual VUV-absorption cross section for 
each feature in the spectra. Sections ~\ref{Astro} and ~\ref{Conclu} summarize the astrophysical implications and the 
conclusions.

\section{Experimental protocol}
\label{Expe}

The measurements were conducted using the Interstellar Astrochemistry Chamber (ISAC). This set-up and the 
standard experimental protocol were described in Mu\~noz Caro et al. \cite{Caro1}. The specific detail on the VUV-measurements of absorption cross 
sections of ice are provided in Paper I. ISAC mainly consists of an ultra-high-vacuum (UHV) 
chamber, with pressure typically in the range of P = 3.0-4.0 $\times$ 10$^{-11}$ mbar, where an ice layer made by deposition of a gas species 
onto a cold finger at 8 K, achieved by means of a closed-cycle helium cryostat, can be UV-irradiated. The evolution of the solid sample 
was monitored with \emph{in situ} transmittance FTIR spectroscopy and VUV-spectroscopy. The chemical components used for the 
experiments described in this paper were CO$_2$(gas), Praxair 99.998\%; CH$_4$(gas), Praxair 99.999\%; N$_2$(gas), Praxair 99.999\%; and O$_{2}$(gas), Praxair 99.8\%. 
The deposited ice layer was photoprocessed with a microwave-discharged hydrogen flow 
lamp (MDHL), from Opthos Instruments. The source has an UV-flux of $\approx 2 \times 10^{14}$ cm$^{-2}$ s$^{-1}$ at the sample 
position, measured by CO$_{2}$ $\to$ CO actinometry, see Mu\~noz Caro et al. \cite{Caro1}. The Evenson cavity of the lamp is refrigerated 
with air. The VUV-spectrum is measured routinely {\em in situ} during the irradiation experiments with the use of a McPherson 0.2-meter 
focal length VUV monochromator (model 234/302) with a photomultiplier tube (PMT) detector equipped with a sodium salicylate window, 
optimized to operate from 100-500 nm (11.27-2.47 eV), with a resolution of 0.4 nm. The characterization of the MDHL spectrum was previously reported 
(Chen et al. 2010, Paper I) and will be discussed in more detail by Chen et al. \cite{Asper2}. For more details and a 
scheme of the experimental set-up we refer to Paper I.

\section{VUV-absorption cross section of interstellar ice analogs}
\label{VUV}

VUV-absorption spectra of pure ices composed of CH$_{4}$, CO$_{2}$, N$_{2}$, and O$_{2}$ have been recorded. 
The column density of the ice sample was measured by FTIR in 
transmittance. The VUV-spectrum and the column density of the ice were therefore monitored in a single experiment for 
the same ice sample. This improvement allowed us to estimate the VUV-absorption cross section of the ice more accurately. The column 
density of the deposited ice obtained by FTIR was calculated according to the formula
\begin{equation}
N= \frac{1}{\mathcal{A}} \int_{band} \tau_{\nu}d{\nu}
\label{1}
\end{equation}
where $N$ is the column density of the ice, $\tau_{\nu}$ the optical depth of the IR band, $d\nu$ the wavenumber differential, 
in cm$^{-1}$, and $\mathcal{A}$ is the band strength in cm molecule$^{-1}$, see Table \ref{table1}. The integrated absorbance is equal to 
0.43 $\times$ $\tau$, where $\tau$ is the integrated optical depth of the IR band. The VUV-absorption cross section was estimated according to the Beer-Lambert law
\begin{eqnarray}
I_t(\lambda) &=& I_0(\lambda) {e}^{-\sigma(\lambda) N} \nonumber \\
\sigma(\lambda) &=& - \frac{1}{N} \ln \left( \frac{I_t(\lambda)}{I_0(\lambda)} \right) \nonumber \\ 
\text{with} \quad N &\approx& \frac{N_i + N_f}{2},
\label{2}
\end{eqnarray}
where I$_{t}(\lambda)$ is the transmitted intensity for a given wavelength $\lambda$, I$_{0}(\lambda)$ the incident intensity, $\sigma$ is the cross section in cm$^{2}$, and 
$N$ is the average ice column density before ($N_i$) and after ($N_f$) exposure to VUV during VUV-spectral acquisition, in cm$^{-2}$. We decided to take this average value of 
$N$ because the ice sample thickness decreases during exposure to VUV photons. If instead of this average value, $N_i$ would be used as $N$ in Eq. 2, this would lead to a lower 
limit of the VUV-absorption cross section. It is important to notice that the total VUV-flux value emitted by the lamp does not affect the VUV-absorption spectrum of the ice sample, 
since it is obtained by subtraction of two spectra to obtain the absorbance in the VUV. 

Several measurements for different values of the initial ice column density were performed to improve the spectroscopy. 
Table ~\ref{table1} provides the infrared-band positions and band strengths of CH$_{4}$ and CO$_{2}$ used to estimate the column density. 
Solid N$_{2}$ and O$_2$ do not display absorption features in the mid-infrared, therefore their column densities were measured using the expression
\begin{equation}
N = \frac{N_A \; \rho_{i} \; d_i}{m_{i}},
\label{3}
\end{equation}
where $N_A$ is the Avogadro constant (6.022 $\times$ 10$^{23}$ mol$^{-1}$), $\rho_{i}$ is the density of the ice in g cm$^{-3}$, see Table~\ref{table1}, m$_i$ is the molar mass of 
the species in g mol$^{-1}$, and d$_i$ is the ice thickness in cm. The latter was estimated following the classical interfringe relation 
\begin{equation}
d_i= \frac{1}{2 n_i \Delta \nu},
\label{4}
\end{equation}
where $n_i$ is the refractive index of the ice at deposition temperature, and $\Delta \nu$ is the wavenumber difference between two adjacent maxima or minima of the fringes 
observed in the infrared spectrum of the ice. These interference fringes are due to multiple reflections of light within the sample.

\begin{table}[hb]
\centering
\caption{Infrared-band positions, infrared-band strengths ($\mathcal{A}$), column density ($N$) in 10$^{15}$ molec./cm$^{2}$, 
refractive index ($n_i$), and the density ($\rho_{i}$) of the samples used in this work.}
\tiny
\begin{tabular}{cccccc}
\hline
\hline
Species&Position&$\mathcal{A}$&$N$&$n_i$&$\rho$\\
&[cm$^{-1}$]&[cm/molec.$^{1}$]&[$\times$ 10$^{15}$ molec./cm$^{2}$]&&[g/cm$^{3}$]\\
\hline
CH$_{4}$&2343&7.6$^{+0.1}_{-0.1}$ $\times10^{-17 \, a}$&146$^{+12}_{-29}$&1.30$^{c}$&0.47$^{c}$\\
CO$_{2}$&1301&6.4$^{+0.1}_{-0.1}$ $\times10^{-18 \, b}$&81$^{+7}_{-11}$&1.21$^{c}$&0.88$^{c}$\\
N$_{2}$&--&--&4774&1.21$^{c}$&0.94$^{c}$\\
O$_{2}$&--&--&60$^{+5}_{-11}$&1.32$^{d}$&1.54$^{d}$\\
\hline
\end{tabular}
\\
{\small $^a$d'Hendecourt \& Allamandola 1986, $^b$Yamada \& Person 1964, $^c$Satorre et al. 2008, $^d$Fulvio et al. 2009}\\
\label{table1}
\end{table}

Solid O$_2$ was deposited at a temperature of 8 K. We used the Fulvio et al. \cite{Fulvio} values of $n_i$ and $\rho$ at 16 K as an approximation. 
Error values for the column density in Table ~\ref{table1} result mainly from the selection of the baseline for integration of the IR absorption band and the decrease of the 
ice column density due to VUV-irradiation during spectral acquisition. The band strengths were adapted from the literature and their error estimates are no more than 10 \% of the 
reported values (Richey \& Gerakines 2012). The errors in the column density determined by IR spectroscopy were 38\%, 22\%, and 35\% for solid CH$_{4}$, CO$_{2}$, 
and O$_{2}$. N$_2$ and O$_2$ do not present any IR feature, but photoprocessed O$_2$ ice during VUV-spectral acquisition produces O$_3$ very readily and the loss of 
O$_2$ can be calculated. On the other hand, N$_2$ ice does not lead to photoproducts and the error in the column density could not be measured. The 
VUV-absorption cross-section errors result from the error values of the column density determination and the MDHL, photomultiplier tube (PMT), and multimeter stabilities, 
about 6\%, using the expression
\begin{equation}
\delta(N) =  \sqrt{\frac{\delta_i^2 + \delta_j^2 + \delta_k^2 + ... + \delta_n^2}{n-1}} .
\end{equation}

The VUV-absorption cross-section spectra of CH$_{4}$, CO$_{2}$, N$_{2}$, and O$_{2}$ ices were fitted with Gaussian profiles using the band positions reported in the literature 
(Mason et al. 2006; Lu et al. 2008; Wu et al. 2012) as a starting point, see the red dotted and dashed-dotted traces in Figs.~\ref{CH4},~\ref{CO2},~\ref{N2}, and ~\ref{O2}. Table~\ref{tableGauss} summarizes 
the Gaussian profile parameters used to fit the spectra of these ices, deposited at 8 K. Gaussian fits of the reported molecules were made with an in-house IDL code. The fits 
reproduce the VUV-absorption cross-section spectra well. 

\begin{table}[ht!]
\centering
\caption{Parameter values used to fit the spectra of Gaussian profiles of the different molecular ices deposited at 8 K. }
\tiny
\begin{tabular}{cccc}
\hline
\hline
Molecule&Center&FWHM&Area\\
&[nm]&[nm]&[$\times$ 10$^{-17}$ cm$^{2}$ nm]\\
\hline
&&&\\
CH$_{4}$&102.0&21.2&90.2\\
&126.0&10.6&15.6\\
&140.0&7.1&0.6\\
\hline
&&&\\
CO$_{2}$&115.3&4.2&1.8\\
&126.4&9.9&2.1\\
\hline
&&&\\
N$_{2}$&115.1&1.06&0.79\\
&116.7&1.06&1.47\\
&118.4&1.06&1.69\\
&120.6&1.06&2.78\\
&122.6&1.06&1.12\\
&123.0&1.59&0.26\\
&124.7&0.74&2.92\\
&125.2&1.59&0.37\\
&127.0&0.74&2.34\\
&127.6&0.94&1.09\\
&129.4&0.71&2.20\\
&130.2&0.71&2.20\\
&131.9&0.71&0.64\\
&132.8&0.80&1.38\\
&134.6&1.06&0.26\\
&135.8&0.80&3.97\\
&137.4&2.12&0.19\\
&138.8&0.94&2.85\\
&142.2&0.80&2.96\\
&145.4&1.06&1.53\\
\hline
&&&\\
O$_{2}$&140.7&25.20&21.46\\
&161.0&23.55&4.01\\
\hline
\end{tabular}
\label{tableGauss}
\end{table}

The main emission peaks of the MDHL occur at 121.6 nm (Lyman-$\alpha$), 157.8 nm, and 160.8 nm (molecular H$_2$ bands). These peaks are thus also present in the secondary VUV photon 
spectrum generated by cosmic rays in dense interstellar clouds and circumstellar regions where molecular hydrogen is abundant (Gredel et al. 1989). For this reason, the 
absorption cross-section values measured at these wavelengths are provided for each molecule in the following sections. We also present an average value of the VUV-absorption cross 
section (a simple arithmetic mean in the spectral range indicated for each species) and an integrated VUV-absorption cross section (an integration of the form $\sigma_{INT} = \int_{\lambda_i}^{\lambda_f} \sigma_{\lambda} d\lambda$) 
in the same spectral range. The same measurements were performed using the gas-phase raw data, adapted from other works, for each species.

\subsection{Solid methane}

The ground state of CH$_{4}$ is \~X$^{1}$A$_{1}$ and its bond energy is E$_{b}$(H--CH$_{3}$) = 4.5 eV (Okabe 1978). 

\begin{figure}[ht!]
\centering
\includegraphics[width=\columnwidth]{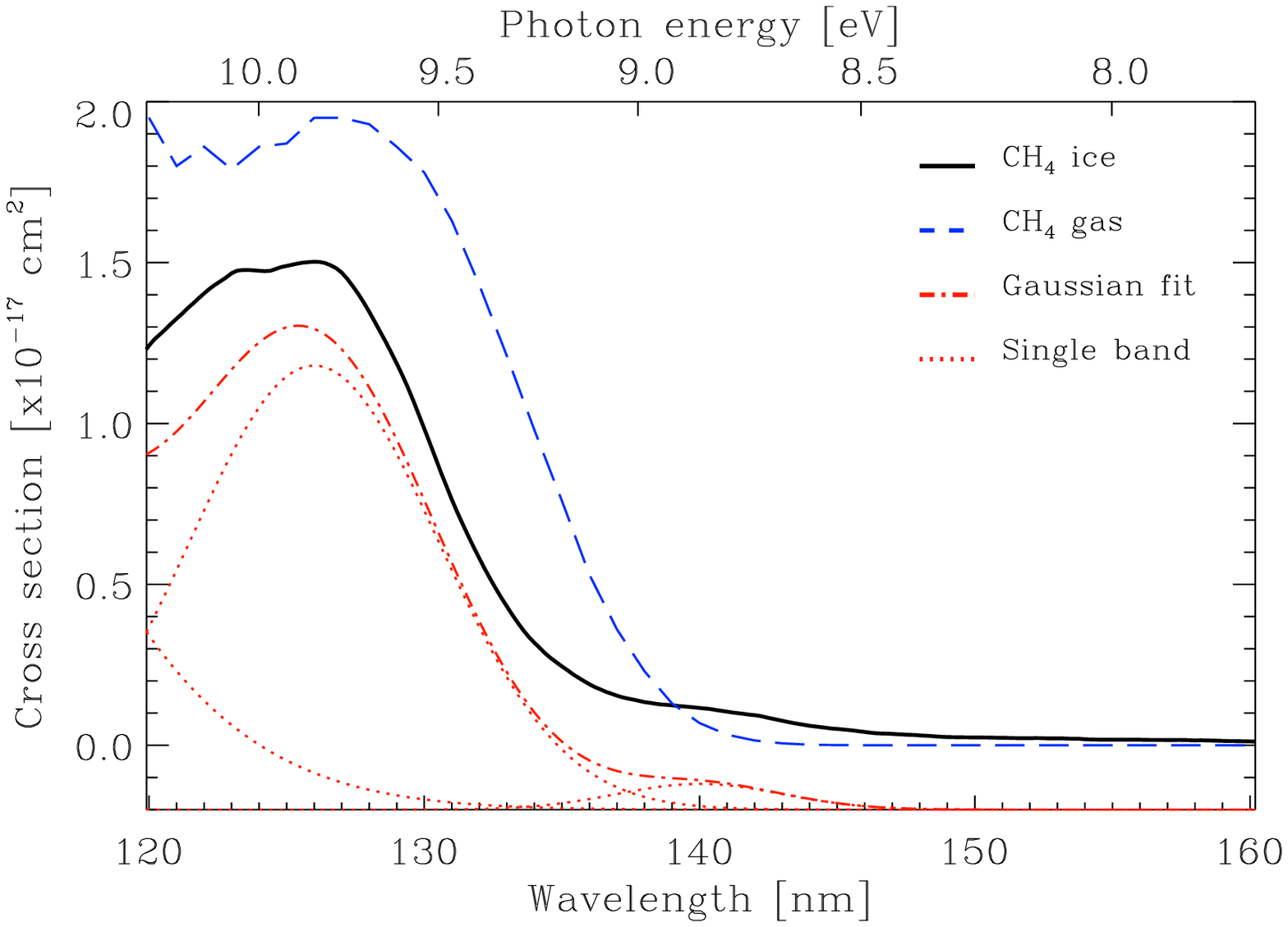}
\caption{VUV-absorption cross section as a function of photon wavelength (bottom X-axis) and VUV-photon energy (top X-axis) of CH$_{4}$ 
ice deposited at 8 K, black solid trace. The blue dashed trace is the VUV-absorption cross-section spectrum of gas phase CH$_{4}$ adapted from Lee et al. 
\cite{Lee}. The fit of the solid-phase spectrum, red dotted (single bands, see Table \ref{tableGauss}) and dashed-dotted trace, is the sum of three Gaussians. It has been offset for clarity.}
\label{CH4}
\end{figure}

Fig.~\ref{CH4} shows the VUV-absorption cross section of CH$_{4}$ as a function of the wavelength and photon energy. The high absorption of CH$_4$ ice around 
and below 120 nm, and the absorption of MgF$_2$ in the same spectral region only allowed spectroscopic measurements starting from 120 nm instead of 113 nm. As in 
Wu et al. \cite{Wu}, a broad absorption band extending to 137 nm (9.05 eV) and centered on 124 nm 
(10.0 eV) was observed. This feature is attributed to the 1t$_{2}$-3s (D$_{2d}$) Rydberg transition. 
We detected a small bump near 140 nm (8.85 eV), which was not observed by Wu et al. \cite{Wu} in the solid phase or by Lee et al. \cite{Lee}
in the gas phase. This feature may be produced by the VUV-absorption of a photoproduct. This phenomenon is more intense in the case of CO$_2$ studied 
in detail in Sect.~\ref{secco2}. 
CH$_{4}$ ice presents high absorption at Ly-$\alpha$ (121.6 nm) and almost no absorption at Lyman band system wavelengths (132-165 nm). 

The average VUV-absorption cross section has a value of 5.7$^{+0.5}_{-1.1}$ $\times$ 10$^{-18}$ cm$^{2}$ in the 120-150 nm (10.33-8.26 eV). 
The total integrated VUV-absorption cross section has a value of 2.0$^{+0.1}_{-0.4}$ $\times$ 10$^{-16}$ cm$^{2}$ nm (1.5$^{+0.1}_{-0.3}$ $\times$ 10$^{-17}$ cm$^{2}$ 
eV) in the same spectral region. The VUV-absorption cross sections of CH$_{4}$ ice at 121.6 nm is 1.4$^{+0.1}_{-0.3}$
$\times$ 10$^{-17}$ cm$^{2}$, that is, at the Ly-$\alpha$ position of atomic hydrogen and there is no observable VUV-absorption in the position of the molecular 
hydrogen emission bands at 157.8 and 160.8 nm. Gas-phase data from Lee et al. \cite{Lee} were used for comparison with our solid-phase data, see Fig.~\ref{CH4}, 
following Wu et al. \cite{Wu}. The solid CH$_{4}$ absorption is shifted to shorter wavelength with respect to the gas data. The VUV-absorption cross section of 
CH$_{4}$ in the gas phase has an average value of 8.2 $\times$ 10$^{-18}$ cm$^{2}$ in the 120-150 nm range. 
CH$_{4}$ gas data were integrated in the same range, giving a value of 2.9 $\times$ 10$^{-16}$ cm$^{2}$ nm 
(2.09 $\times$ 10$^{-17}$ cm$^{2}$ eV), which is higher than the solid CH$_{4}$ value. The VUV-absorption cross section of
CH$_{4}$ gas at 121.6 nm is 1.8 $\times$ 10$^{-17}$ cm$^{2}$.

\subsection{Solid carbon dioxide}
\label{secco2}

The ground state and bond energy of CO$_{2}$ are X$^{1}\Sigma_{g}^{+}$ and E$_{b}$(O--CO) = 5.5 eV (Okabe 1978). 

\begin{figure}[ht!]
\centering
\includegraphics[width=\columnwidth]{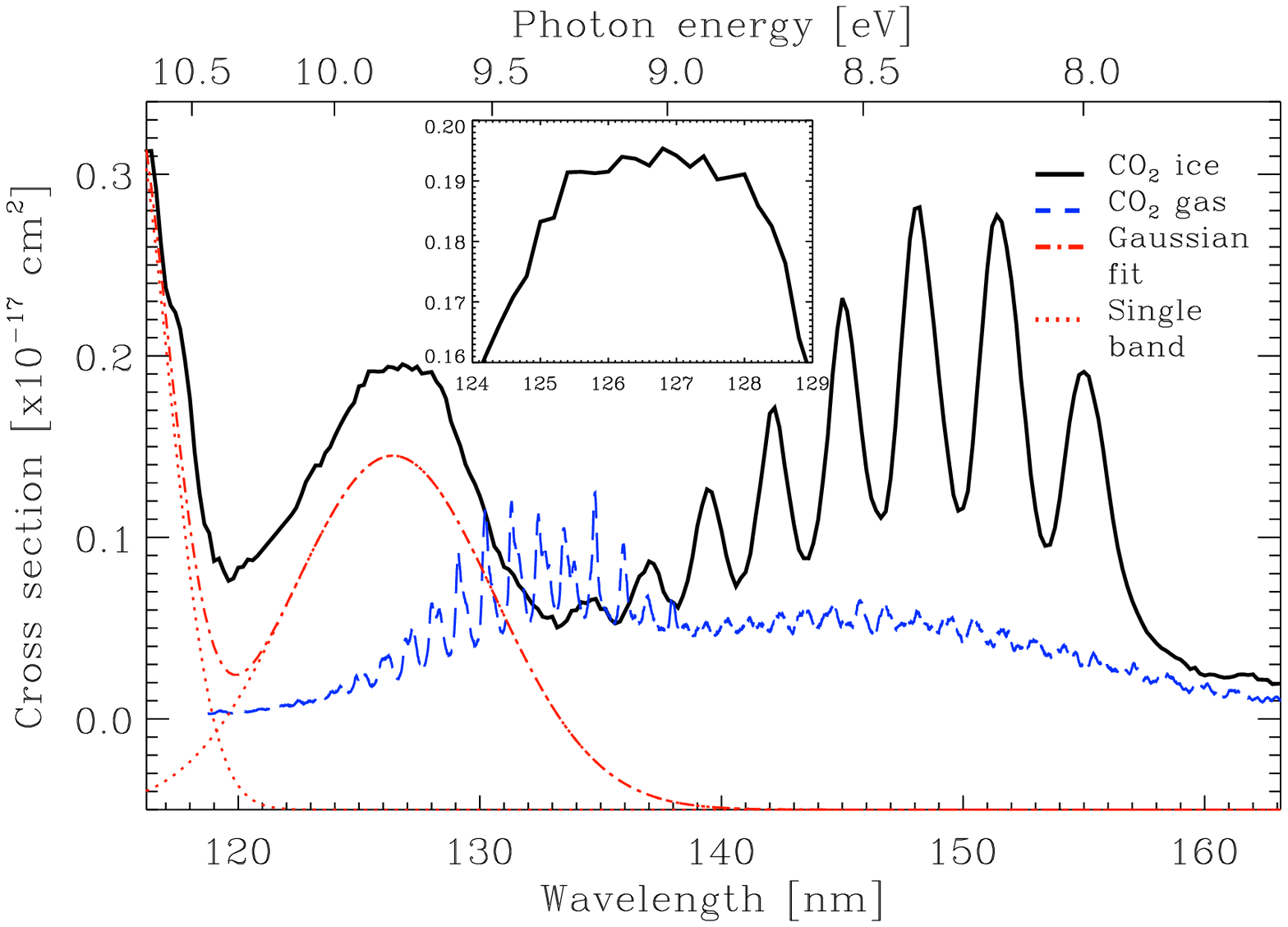}
\caption{VUV-absorption cross section as a function of photon wavelength (bottom X-axis) and VUV-photon energy (top X-axis) of CO$_{2}$ 
ice deposited at 8 K, black solid trace. The blue dashed trace is the VUV-absorption cross-section spectrum of gas phase CO$_{2}$ adapted from Yoshino et 
al. \cite{Yoshino}. The fit, solid red dotted (single bands, see Table \ref{tableGauss}) and dashed-dotted trace, is the sum of two Gaussians. The inset figure is a CO$_{2}$ 
VUV-absorption cross-section close-up in the 124-129 nm range.}
\label{CO2}
\end{figure}

\begin{figure}[ht!]
\centering
\includegraphics[width=\columnwidth]{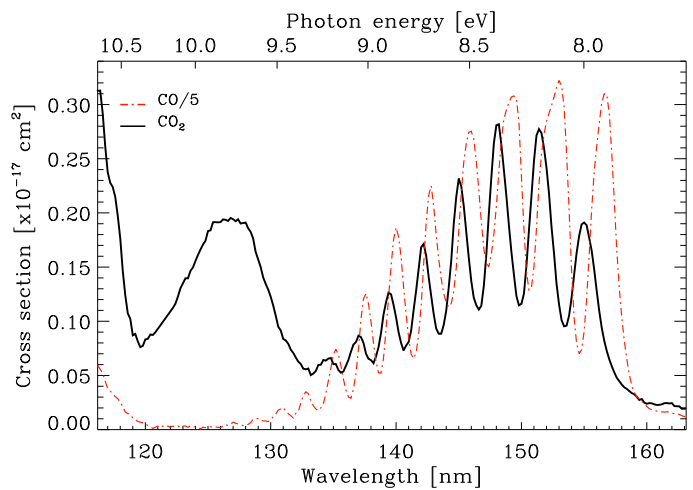}
\caption{Comparison between the VUV-absorption cross-section spectra of CO (red dashed-dotted trace) and CO$_{2}$ (black solid trace, the CO$_2$ ice is mixed 
with CO formed during the spectral acquisition). CO transitions in a CO$_{2}$ matrix are shifted to shorter wavelengths with respect to pure CO ice transitions.}
\label{CO2CO}
\end{figure}

The VUV-absorption cross section of CO$_{2}$ as a function of the wavelength and photon energy is shown in Fig.~\ref{CO2}. This spectrum is 
very different from those reported in Lu et al. \cite{Lu2}, Mason et al. \cite{Mason}, and Monahan \& Walker \cite{Monahan}. 
Mason et al. \cite{Mason} observed two broad bands centered on 8.8 and 9.9 eV assigned to the 
$^{1}\Delta_{u}\leftarrow ^{1}\Sigma^{+}_{g}$ and $^{1}\Pi_{g} \leftarrow ^{1}\Sigma^{+}_{g}$ transitions. 
Lu et al. \cite{Lu2} and Monahan \& Walker \cite{Monahan} reported a broad band centered on 9.8 eV. 
These authors found a vibrational structure in this band with similar positions. The intense absorption feature 
in the 107-120 nm (11.53-10.33 eV) region reported by Lu et al. \cite{Lu2} is partially observed in our data. Two broad bands 
in the 120-133 nm (10.33-9.32 eV) and 133-163 nm (9.32-7.60 eV) regions were also reported by Lu et al. \cite{Lu2}, which are 
observed in our spectrum as well. A vibrational structure was detected in the 120-133 nm band, see Fig.~\ref{CO2} inlet. This vibrational structure is very 
faint in our spectrum, but 5 out of the 12 bands reported by Mason et al. \cite{Mason} are observable, see Table ~\ref{TableCO2}. 
In the 133-167 nm wavelength range of Fig.~\ref{CO2}, the spectrum presents a CO-like VUV-absorption. Indeed, the VUV-light cone of our 
MDHL source processes the entire volume of the CO$_{2}$ ice and efficiently leads to the formation of CO by photodissociation of CO$_{2}$ molecules. 
After 9 minutes, corresponding to the collection time of a spectrum, CO is present with a column density that is 22 \% of the deposited CO$_{2}$ 
column density, estimated from integration of the IR bands, data not shown. This is enough, for a molecule like CO with a VUV-absorption cross section seven times larger  
than CO$_{2}$, to appear in the VUV-spectrum of CO$_{2}$ ice. The measured VUV-spectrum therefore corresponds to a mixture of CO$_{2}$ and CO, but 
because of the interaction of CO molecules with the CO$_{2}$ ice matrix, the CO features are shifted to shorter wavelengths than those of pure CO ice 
reported, for instance, in Paper I. This is shown in Fig.~\ref{CO2CO}. Table ~\ref{TableCO2CO} summarizes the peak positions of the detected CO 
features in the CO$_{2}$ ice and those reported for pure CO gas and ice. The $\Delta \lambda$- column in Table ~\ref{TableCO2CO} represents 
a redshift with respect to pure CO gas and  $\Delta \lambda$+ represents a blueshift with respect to the pure CO ice. These shifts gradually 
increase at higher wavelengths. The Davydov splitting is not observed in the features of CO in a CO$_{2}$ matrix where the (2,0) transition 
is the strongest (as in pure CO gas; see, e.g., Paper I) in contrast to the VUV-absorption spectrum of pure CO ice, where the (1,0) transition dominates. 
The band positions of CO$_2$ are not significantly affected by the presence of CO because the broad band centered on 9.8 eV (126.5 nm) and the
vibrational structure in the 124-129 nm range of CO$_2$ present no detectable perturbations compared with other works with a much lower CO$_2$ photoproduction during 
spectral acquisition, and the band positions also agree fairly well with Lu et al. \cite{Lu2}, Mason et al. \cite{Mason}, and Monahan \& Walker \cite{Monahan}. 
The resulting fit and Gaussians involved are displayed in Fig.~\ref{CO2} as a red dashed-dotted trace. The 142.5 nm (8.70 eV) band 
reported by Lu et al. \cite{Lu2} was not fitted because of the overlap with the features of photoproduced CO in our experiment. The exact peak position 
of the feature at wavelengths shorter than 120 nm could not be confirmed because it is beyond our spectral range.

An upper limit for the average and the total integrated VUV-absorption cross section was calculated by subtracting CO, they are 
6.7$^{+0.5}_{-0.9}$ $\times$ 10$^{-19}$ cm$^{2}$ and 2.6$^{+0.2}_{-0.3}$ $\times$ 10$^{-17}$ cm$^{2}$ nm (1.9$^{+0.1}_{-0.2}$ $\times$ 10$^{-18}$ cm$^{2}$ eV) in the wavelength 
range of 120-160 nm (10.33-7.74 eV). The VUV-absorption cross section of CO$_{2}$ 
ice at 121.6 nm is 1.0$^{+0.1}_{-0.2}$ $\times$ 10$^{-18}$ cm$^{2}$. The VUV-absorption cross section at the Lyman band system range (132-165 nm) was not measured becuase of the presence 
of CO. Gas-phase data from Yoshino et al. \cite{Yoshino} were used for comparison with 
the solid phase data, see blue trace in Fig.~\ref{CO2}. Mason et al. \cite{Mason} reported similar values for the gas- and solid-phase 
VUV-absorption cross sections, but using Yoshino et al. \cite{Yoshino} gas phase data, and our solid phase data we can clearly observe that the VUV-absorption cross section of 
CO$_{2}$ gas is lower than the solid-phase value, see Fig.~\ref{CO2}. The Mason et al. \cite{Mason} solid-phase spectra displays a maximum at 9.9 eV with a 
VUV-absorption cross section of 1.2 $\times$ 10$^{-18}$ cm$^{2}$, lower than the 1.9$^{+0.1}_{-0.2}$ $\times$ 10$^{-18}$ cm$^{2}$ value deduced from our data. 
The VUV-absorption cross section of CO$_{2}$ in the gas-phase data from Yoshino et al. \cite{Yoshino} has an average value of 3.3 $\times$ 10$^{-19}$ cm$^{2}$ 120-160 nm range. 
CO$_{2}$ gas data were integrated in the spectral range, giving a value of 1.5 $\times$ 10$^{-17}$ cm$^{2}$ nm 
(8.2 $\times$ 10$^{-19}$ cm$^{2}$ eV), which is lower than the value for solid CO$_{2}$ in the same range 
(2.6$^{+0.2}_{-0.3}$ $\times$ 10$^{-17}$ cm$^{2}$ nm). The VUV-absorption cross section of CO$_{2}$ gas at 121.6 nm is very low, 6.3 $\times$ 10$^{-20}$ cm$^{2}$. 

It should be noted that, despite the photoproduced CO in our VUV-spectra of CO$_2$, the VUV-absorption cross sections are similar to 
previously reported values. The data presented in this section provide some evidence on the CO$_2$:CO 
mixture effects in the VUV-spectrum. The VUV-spectroscopy of ice mixtures has been poorly studied (Wu et al. 2012), but it is essential for understanding the absorption 
of UV-photons in icy grain mantles.

\begin{table}[ht!]
\centering
\caption{Transitions observed in the VUV-absorption cross-section spectrum of CO$_{2}$ ice in the 120-133 nm region. The positions agree 
fairly well with Mason et al. \cite{Mason} and Monahan \& Walker \cite{Monahan}.}
\begin{tabular}{cccc}
\hline
\hline
\multicolumn{2}{c}{This work}&Mason et al. &Monahan \& Walker\\
&&\cite{Mason}&\cite{Monahan}\\
\multicolumn{2}{c}{[nm] $\;$ [eV]}&[eV]&[eV]\\
\hline
125.0&9.91&9.93&9.90\\
%125.4&9.88&9.85&9.82\\
126.4&9.82&9.85&9.82\\
126.8&9.77&9.77&9.74\\
127.4&9.73&9.70&9.67\\
128.0&9.68&9.62&9.59\\
\hline
\end{tabular}
\label{TableCO2}
\end{table}

\begin{table}[ht!]
\centering
\caption{Transitions of pure CO in gas and ice phases, and in a CO$_{2}$ matrix with a CO abundance of 22 \%. CO gas transitions 
are shifted to shorter wavelengths with respect to CO in a CO$_{2}$ matrix (column $\Delta \lambda$-), which in turn are shifted to shorter wavelengths 
with respect to pure CO ice transitions (column $\Delta \lambda$+).}
\tiny
\begin{tabular}{cccc}
\hline
\hline
&Lu et al. \cite{Lu}&This work&Paper I\\
($\nu$',$\nu$'')&Pure CO gas&CO in CO$_{2}$ matrix&Pure CO ice\\
&[nm]&$\Delta \lambda$- $\;$ [nm] $\;$ $\Delta \lambda$+&[nm]\\
\hline
7,0&134.7&0.1 $\;$ 134.8 $\;$ 0.4&135.2\\
6,0&136.9&0.1 $\;$ 137.0 $\;$ 0.6&137.6\\
5,0&139.3&0.1 $\;$ 139.4 $\;$ 0.6&140.0\\
4,0&142.1&0.1 $\;$ 142.2 $\;$ 0.6&142.8\\
3,0&144.8&0.2 $\;$ 145.0 $\;$ 1.0&146.0\\
2,0&147.9&0.3 $\;$ 148.2 $\;$ 1.2&149.4\\
1,0&151.1&0.3 $\;$ 151.4 $\;$ 1.6&153.0\\
0,0&154.4&0.6 $\;$ 155.0 $\;$ 1.6&156.6\\
\hline
\end{tabular}
\label{TableCO2CO}
\end{table}

%\begin{figure}[ht!]
%\centering
%\includegraphics[width=\columnwidth]{UVlamp/Cross_CO2_UV_inlet.ps}
%%\includegraphics[width=7cm]{UVlamp/Cross_CO2_UV.ps}
%\caption{CO$_{2}$ VUV-absorption cross section close-up. It shows 5 of the 12 features reported by Mason et al. \cite{Mason}. DISCUTIR LA FALTA DE PICOS}
%\label{CO22}
%\end{figure}

\subsection{Solid nitrogen}
\label{SN2}

The ground state and bond energy of N$_{2}$ are X$^{1}\Sigma_{g}^{+}$ and E$_{b}$(N--N) = 9.8 eV (Okabe 1978). 

\begin{figure}[ht!]
\centering
\includegraphics[width=\columnwidth]{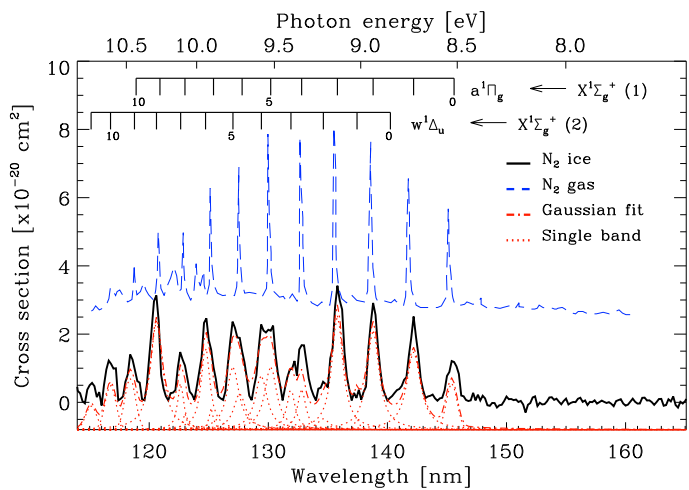}
\caption{VUV-absorption cross section as a function of photon wavelength (bottom X-axis) and VUV-photon energy (top X-axis) of N$_{2}$ 
ice deposited at 8 K, black solid trace. The blue dashed trace is the VUV-absorption cross-section spectrum of gas phase N$_{2}$ adapted from Mason et al. \cite{Mason}. 
The fit, red dotted (single bands, see Table \ref{tableGauss}) and dashed-dotted trace, is the sum of twenty Gaussian profiles. It has been offset for clarity.}
\label{N2}
\end{figure}

Owing to its very low VUV-absorption cross section, a deposition of 2.36 $\mu$m, nearly 4.7 $\times$ 10$^{18}$ molecules cm$^{-2}$, 
for N$_{2}$ ice was required to detect the absorption features. 
The VUV-absorption cross section as a function of wavelength and photon energy is shown in Fig.~\ref{N2}. It presents a 
vibrational structure in the 114-147 nm (10.87-8.43 eV) region. Two systems (attributed to a$^{1}\Pi_{g}$ 
$\leftarrow$ X$^{1}\Sigma_{g}^{+}$ and w$^{1}\Delta_{u}$ $\leftarrow$ X$^{1}\Sigma_{g}^{+}$) can be observed. These systems have been 
reported by Boursey et al. \cite{Boursey}, Mason et al. \cite{Mason}, and Wu et al. \cite{Wu}. In agreement with Boursey et al. \cite{Boursey} 
and Wu et al. \cite{Wu}, we did not observe the two broad continua between 133-163 nm (9.30-7.60 eV) and 
113-133 nm (9.30-11.0 eV) that Mason et al. \cite{Mason} reported. The VUV-absorption cross-section spectrum of N$_{2}$ is not 
as well resolved as in other works, but this probably does not affect the VUV-absorption cross-section scale we provide because no integration 
of the band area is involved. Table ~\ref{TableN2} summarizes the peak position and the band area integration of all the features presented. 
Even though our UV-lamp flux decreases below 120 nm, three features are observed in that spectrum range. 

The average VUV-absorption cross section has a value of 7.0 $\times$ 10$^{-21}$ cm$^{2}$ in the 114.6-146.8 nm (10.82-8.44 eV) range. 
The total integrated VUV-absorption cross section has a value of 2.3 $\times$ 10$^{-19}$ cm$^{2}$ nm (1.8 $\times$ 10$^{-20}$ cm$^{2}$ 
eV) in the same spectral region. The latter value is indeed affected by the low resolution of our VUV-absorption spectrum. A spectrum with better resolved bands would result in a 
lower value. Comparing our value of VUV-absorption cross section with data from Wu et al. \cite{Wu}, we estimated an increase of 34\% in the total integrated VUV-absorption cross-section 
value, therefore this integrated value would be $\sim$ 1.5 $\times$ 10$^{-19}$ cm$^{2}$ nm. The VUV-absorption cross section at Ly-$\alpha$ (121.6 nm) is very low; we 
estimated an upper limit value of 1.0 $\times$ 10$^{-21}$ cm$^{2}$. There is no 
observable VUV-absorption at molecular hydrogen-band wavelengths (157.8 and 160.8 nm). Gas-phase VUV-absorption data adapted from Mason et al. \cite{Mason} are 
plotted in Fig.~\ref{N2} as a blue dashed trace. The gas data of Mason et al. \cite{Mason} are not in cross section units, this is why we did not calculate the average and the 
total integrated VUV-absorption cross sections.

\begin{table}
\centering
\caption{Transitions observed in the VUV-absorption cross-section spectrum of N$_{2}$. Columns (1) and (2) values are represented in 
Fig.~\ref{N2}.}
\tiny
\begin{tabular}{ccccc}
\hline
\hline
($\nu$',$\nu$'')&\multicolumn{2}{c}{Position}&\multicolumn{2}{c}{Area}\\
(1)$\quad$ $\;$(2)&[nm]&[eV]&[cm$^{2}$ nm]&[cm$^{2}$ eV]\\
\hline
\hspace{0.5cm}11,0&115.2&10.76&7.0$\times10^{-22}$&8.1$\times10^{-23}$\\
\hspace{0.5cm}10,0&116.6&10.63&2.7$\times10^{-21}$&2.5$\times10^{-22}$\\
\hspace{-0.26cm} 10,0+9,0&118.4&10.47&1.1$\times10^{-20}$&1.0$\times10^{-21}$\\
9,0+8,0&120.6&10.28&2.2$\times10^{-20}$&1.9$\times10^{-21}$\\
8,0+7,0&122.6&10.11&9.0$\times10^{-21}$&6.8$\times10^{-22}$\\
7,0+6,0&124.8&9.93&1.9$\times10^{-20}$&1.6$\times10^{-21}$\\
6,0+5,0&127.0&9.76&2.5$\times10^{-20}$&1.9$\times10^{-21}$\\
5,0+4,0&129.4&9.58&2.6$\times10^{-20}$&1.9$\times10^{-21}$\\
4,0+3,0&133.0&9.32&1.6$\times10^{-20}$&1.2$\times10^{-21}$\\
3,0+2,0&135.8&9.12&2.3$\times10^{-20}$&1.6$\times10^{-21}$\\
2,0+1,0&138.8&8.93&2.2$\times10^{-20}$&1.5$\times10^{-21}$\\
1,0+0,0&142.2&8.71&2.0$\times10^{-20}$&1.2$\times10^{-21}$\\
0,0$\quad$ $\;$ $\;$&145.4&8.52&1.2$\times10^{-20}$&7.2$\times10^{-22}$\\
\hline
\end{tabular}
\label{TableN2}
\end{table}

\subsection{Solid oxygen}

The ground state of O$_{2}$ is X$^{3}\Sigma_{g}^{-}$ and its bond energy is E$_{b}$(O--O) = 5.1 eV (Okabe 1978). 

\begin{figure}[ht!]
\centering
\includegraphics[width=\columnwidth]{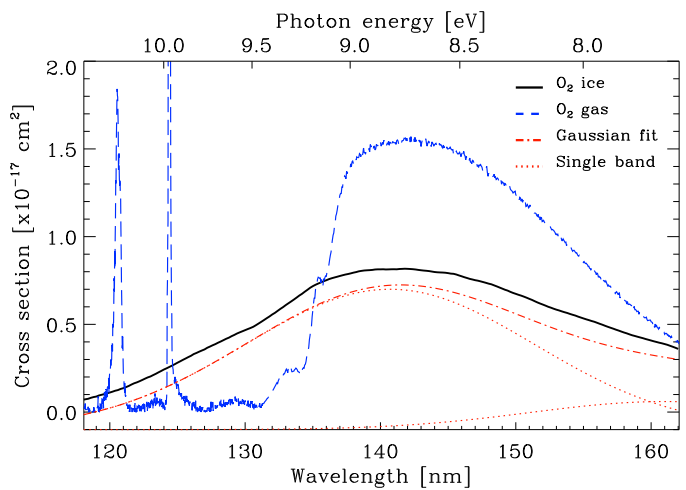}
\caption{VUV-absorption cross section as a function of photon wavelength (bottom X-axis) and VUV-photon energy (top X-axis) of O$_{2}$ 
ice deposited at 8 K, black solid trace. The blue dashed trace is the VUV-absorption cross-section spectrum of gas phase O$_{2}$ adapted from Lu et al. 
\cite{Lu3}. The fit, solid red dotted (single bands, see Table \ref{tableGauss}) and dashed-dotted trace, is the sum of two Gaussians. It has been offset for clarity.}
\label{O2}
\end{figure}

The VUV-absorption cross section as a function of wavelength and photon energy is shown in Fig.~\ref{O2}. The O$_2$ ice thickness estimation based on the 
fringes method, described in Sect.~\ref{VUV}, requires a thick ice of about 60 ML. But the VUV-absorption cross-section measurement only works well for 
thin ices below 200 ML. Therefore, several experiments were performed to ensure that the ice thickness estimation was correct in the case of O$_2$ . Solid O$_{2}$ presents a broad 
band centered on 141 nm (8.79 eV) in the 118-162 nm (10.50-7.65 eV) region (attributed to the \~B$^{3}\Sigma_{u}^{-}$ $\leftarrow$ 
X$^{3}\Sigma_{g}^{-}$ transition, named Schumann-Runge band). It also presents a feature centered on 177 nm (7.0 eV), according to 
Mason et al. \cite{Mason} and Lu et al. \cite{Lu2}, which is beyond our spectral range. The resulting fit and Gaussians involved are represented in Fig.~\ref{O2} 
by a red dashed-dotted trace. Following Lu et al. \cite{Lu2}, there is no obvious requirement for a broad line centered near 161 nm (7.70 eV), but 
without this band the resulting fit in the 150-162 nm range would not be as good. 

The average VUV-absorption cross section has a value of 4.8$^{+0.4}_{-1.0}$ $\times$ 10$^{-18}$ cm$^{2}$ in the 120-162 nm (10.33-7.65 eV) range. 
The total integrated VUV-absorption cross section has a value of 2.4$^{+0.2}_{-0.5}$ $\times$ 10$^{-16}$ cm$^{2}$ nm (1.5$^{+0.1}_{-0.3}$ $\times$ 10$^{-17}$ cm$^{2}$ 
eV) in the same spectral region. The VUV-absorption cross sections of the O$_{2}$ ice at Ly-$\alpha$ (121.6 nm) and 
the molecular hydrogen band emissions (157.8 and 160.8 nm) are 1.4$^{+0.1}_{-0.3}$ $\times$ 10$^{-18}$ cm$^{2}$, 4.6$^{+0.4}_{-0.9}$ $\times$ 10$^{-18}$ cm$^{2}$, and 
3.9$^{+0.3}_{-0.8}$ $\times$ 10$^{-18}$ cm$^{2}$.
Gas-phase data from Lu et al. \cite{Lu3} were used for comparison with our solid phase data, see Fig.~\ref{O2}. Solid- and gas-phase data 
have a broad absorption band, but only the gas-phase spectrum presents discrete transitions. The VUV-absorption cross section of O$_{2}$ 
in the gas phase has an average value of 4.0 $\times$ 10$^{-18}$ cm$^{2}$. The O$_{2}$ gas spectrum was integrated in the 
118-162 nm range, giving a value of 3.3 $\times$ 10$^{-16}$ cm$^{2}$ nm (2.0 $\times$ 10$^{-17}$ cm$^{2}$ eV), which is higher than the solid-O$_{2}$ 
value. The VUV-absorption cross sections of O$_{2}$ gas at 121.6 nm, 157.8 nm and 160.8 nm are 2.5 $\times$ 10$^{-18}$ 
cm$^{2}$, 6.5 $\times$ 10$^{-18}$ cm$^{2}$, and 4.7 $\times$ 10$^{-18}$ cm$^{2}$, which is also higher than the solid-phase 
measurements, with the exception of the Ly-$\alpha$ wavelength (121.6 nm).

\subsection{Comparison between all the species}

\begin{figure}[ht!]
\centering
\includegraphics[width=\columnwidth]{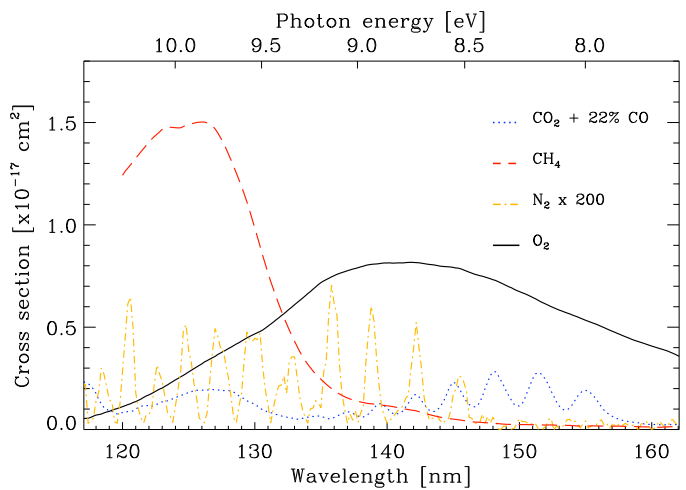}
\caption{VUV-absorption cross section as a function of wavelength (lower X-axis) and photon energy (top X-axis) of all the 
species in the solid phase at 8 K. The CO$_2$ absorption only occurs at wavelengths shorter than 133 nm, 
the absorption at longer wavelengths is due to the photoproduced CO during spectral adquisition.}
\label{todas}
\end{figure}

Fig.~\ref{todas} shows a comparison of the VUV-absorption cross section for all the species represented in the same linear scale. Because of 
its low VUV-absorption cross section, the N$_{2}$ spectrum was multiplied by a factor of 200 for better appreciation. It can be observed that the most highly absorbing molecule 
around the hydrogen Ly-$\alpha$ (121.6 nm) position is CH$_{4}$. Solid O$_{2}$ is most highly absorbing around the Lyman band system. N$_{2}$ gives the lowest 
absorption in both cases. All the species absorb at the Ly-$\alpha$ wavelength, with the exception of N$_{2}$, for which an upper limit was given in Sect.~\ref{SN2}. CH$_{4}$ and 
N$_{2}$ absorb VUV-photons with wavelengths lower than 150 nm, while O$_{2}$ absorbs throughout the scanned range. Table ~\ref{tableInt} compares the VUV-absorption 
cross sections of all the species in the gas and solid phase 
as well as an average VUV-absorption cross section in the 120.8-122.6 nm and 132-162 nm range normalized by the Ly-$\alpha$ and Lyman band system photon flux 
of our MDHL.

\begin{table*}
\centering
\caption{Comparison between the VUV-absorption cross sections in the 120-160 nm range of all the species in the gas and solid phase. 
Total Int. is the total integrated VUV-absorption cross section in this range, Avg. is the average VUV-absorption cross section. 
Ly-$\alpha$ and LBS are the average VUV-absorption cross sections in the 120.8-122.6 nm and 132-162 nm range normalized by the  
photon flux. The last three columns are the VUV-absorption cross sections at wavelengths 121.6 nm (corresponding to the maximum intensity 
of the Ly-$\alpha$ peak), 157.8 and 160.8 nm (main peaks of the Lyman band system). All the VUV-absorption cross-section values in the table need to 
be multiplied by 1 $\times$ 10$^{-18}$ cm$^2$, as indicated below.}
\small
\begin{tabular}{ccccccccc}
\hline
\hline
&Species&Total Int.&Avg.&Ly-$\alpha$&LBS&121.6 nm&157.8 nm&160.8 nm\\
&&[$\times$ 10$^{-16}$ cm$^{2}$ nm]&\multicolumn{6}{c}{[$\times$ 10$^{-18}$ cm$^{2}$]}\\
\hline
\multirow{8}{*}{\rotatebox{90}{Solid phase}}&&&&&&&&\\
&CH$_{4}$&2.0$^{+0.1}_{-0.4}$&5.7$^{+0.5}_{-1.1}$&15&1.5&14$^{+0.1}_{-0.3}$&---&---\\
&&&&&&&&\\
&CO$_{2}$&0.26$^{+0.02}_{-0.03}$&0.67$^{+0.5}_{-0.9}$&1.0&3.6&1.0$^{+0.1}_{-0.2}$&---&---\\
&&&&&&&&\\
&N$_{2}$&0.0023&0.007&0.003&0.001&0.001&---&---\\
&&&&&&&&\\
&O$_{2}$&2.4$^{+0.2}_{-0.5}$&4.8$^{+0.4}_{-1.0}$&1.5&5.4&1.4$^{+0.1}_{-0.3}$&4.6$^{+0.4}_{-0.9}$&3.9$^{+0.3}_{-0.8}$\\
\hline
\hline
\multirow{8}{*}{\rotatebox{90}{Gas phase}}&&&&&&&&\\
&CO$_{2} \quad ^a$&0.15&0.33&---&---&0.063&0.17&0.095\\
&&&&&&&&\\
&CH$_{4} \quad ^b$&2.9&8.2&---&---&18&---&---\\
&&&&&&&&\\
&N$_{2} \quad ^c$&---&---&---&---&---&---&---\\
&&&&&&&&\\
&O$_{2} \quad ^d$&3.3&4.0&---&---&2.5&6.5&4.7\\
\hline
\end{tabular}
\\
{\small Raw data adapted from $^a$ Lee et al. \cite{Lee}, $^b$ Yoshino et al. \cite{Yoshino}, $^c$ Mason et al. \cite{Mason}, and $^d$ Lu et al. \cite{Lu3} were used to calculate 
the reported gas-phase values.}\\
\label{tableInt}
\end{table*}

\section{Astrophysical implications}
\label{Astro}

Interstellar and circumstellar ice mantles are composed of several species such as H$_{2}$O, CO, NH$_{3}$, CO$_{2}$, CH$_{4}$, and CH$_{3}$OH 
(Mumma \& Charnley 2011, and ref. therein). In some cases, these molecular ice components are mixed in the ice mantle.

The VUV-absorption cross-section spectrum of CO$_{2}$ presented in this work has the disadvantage of being mixed with the CO produced by 
the VUV-irradiation of the MDHL used to acquire the VUV-spectroscopy of the species. But on the other hand, our data show that a binary ice mixture containing CO$_{2}$ and 
CO can lead to a mixed VUV-absorption spectrum where mixture effects can be appreciated, using this new way of performing VUV-spectroscopy (cf Wu et al. 2010, 2012), 
where the band profiles and positions are altered to some extent by mixture effects.

As is commonly observed for example in infrared spectra, multicomponent ice mantles most likely display VUV-absorption features that correspond to the various species, where CO is the most 
highly absorbing molecule in the VUV among the common ice components. VUV-irradiation of CO2 ice led to a 22 \% of CO with respect to the starting CO$_2$ column density. CH$_3$OH ice led 
to 16 \% of CO for the same UV-fluence, see Paper I. The two production rates are similar but only the VUV-absorption cross-section spectrum of CO$_2$ ice presents 
CO features. On the other hand, the average VUV-absorption cross section of CH$_3$OH ice is almost ten times larger than the average VUV-absorption cross section of CO$_2$ in 
the 133-160 nm range.

The VUV-spectrum of CO in a CO$_{2}$ ice matrix is different from that of pure CO ice, with shifted positions that occur half-way between the gas 
and solid CO values, see Table  ~\ref{TableCO2CO} and Fig.~\ref{CO2CO}. Therefore, CO in a solid CO$_{2}$ matrix can be treated as a highly porous CO ice mantle. 
Infrared spectroscopy of CO$_2$:CO = 20:1 ice shows a similar trend, the CO band shifts from 2138.6 cm$^{-1}$ (for pure CO ice) to 2139.7 cm$^{-1}$ and the FWHM increases from 
2.5 cm$^{-1}$ (pure CO ice) to 5.8 cm$^{-1}$ (Sandford et al. 1988)

%CH$_4$ is easily dissociated and it will be drastically affected by Ly-$\alpha$ photons. Its VUV-absorption cross section is 
Formation of  N$_{3}$ from N$_{2}$ irradiation with VUV-photons is extremely difficult (Hudson \& Moore 2002). Proton and electron 
bombardment is more effective in the production of the azide radical than VUV-photons (Hudson \& Moore 2002; Jamieson \& Kaiser 2007). 
The likely explanation is the low VUV-absorption cross section of N$_{2}$. Detection of N$_{3}$ in space might be indicate a cosmic-ray field, 
or alternatively VUV-photons with energies higher than Ly-$\alpha$ (10.2 eV) that we were unable to study in this work. The latter 
possibility is very unlikely because much more intense absorption is not expected in the short interstellar radiation field range, from 91.2 to 114 nm, which 
was not measured in the laboratory.

Regarding the two homonuclear molecules studied,  O$_{2}$ and N$_{2}$, the former presents a broad absorption band in the 118-160 nm range with 
a VUV-absorption cross section on the order of 10$^{-18}$ cm$^{-2}$, while N$_{2}$ presents a vibrational structure in the 118-150 nm range with 
a VUV-absorption cross section on the order of 10$^{-20}$ cm$^{-2}$. Using the Pontoppidan \cite{Pontoppidan2} approximation of the abundances of nitrogen and oxygen, 
these two orders of magnitude differences can determine different photochemistry efficiencies for N$_2$ and O$_2$. This supports the hypothesis that a photochemistry richer in O than in N is 
expected in icy grain mantles, because a significant fraction of the N atoms might be locked as solid N$_2$.

The ice penetration depth of photons with a given wavelength, or the equivalent absorbing ice column  density of a species in the solid 
phase, can be calculated from the VUV-absorption cross section following
\begin{equation}
% \hspace{3cm}
N(\lambda)= \frac{- 1}{\sigma(\lambda)} \ln \left( \frac{I_t(\lambda)}{I_0(\lambda)} \right), 
\end{equation}
which can be converted into an ice thickness using equation 3, inserting the volumetric density of the ice provided in Table 1: 
\begin{equation}
d = \frac{N \; m_{i}}{N_A \; \rho_{i}},
\end{equation}
where I$_{t}(\lambda)$ is the transmitted intensity for a given wavelength $\lambda$, I$_{0}(\lambda)$ the incident intensity, $N(\lambda)$ is 
the absorbing column density in cm$^{-2}$, and $\sigma(\lambda)$ is the cross section in cm$^{2}$. Following the estimation for polar ice molecules provided in Paper I, 
Table 7 provides the column density values of nonpolar ice species that correspond to an absorbed flux of 95\% and 99\% using the cross-section 
value at Ly-$\alpha$, the average cross section in the 120-160 nm range, and the maximum cross section in the same range. 

\begin{table}[ht!]
\centering
\caption{Column density of the different ice species corresponding to an absorbed 
photon flux of 95\% and 99\%. Ly-$\alpha$ corresponds to the cross section at the Ly-$\alpha$ wavelength, 121.6 nm; in the case 
of N$_2$, the upper limit in the cross section leads to a lower limit in the absorbing column density. Avg. corresponds to the average 
cross section in the 120-160 nm range. Max. corresponds to the maximum cross section in the same wavelength range.}
\tiny
\begin{tabular}{ccccccc}
\hline
\hline
&\multicolumn{3}{c}{95\% photon absorption}&\multicolumn{3}{c}{99\% photon absorption}\\
Species&Ly-$\alpha$& Avg.& Max.& Ly-$\alpha$& Avg.&Max.\\
&\multicolumn{3}{c}{ ($\times$10$^{17}$ molec./cm$^{2}$)}&\multicolumn{3}{c}{($\times$10$^{17}$ molec./cm$^{2}$)}\\
\hline
CH$_4$&2.1&5.3&1.8&3.3&8.1&2.7\\
CO$_2$&29.3&44.5&15.1&45.1&68.4&23.3\\
N$_2$&29957&4280&881&46052&6579&1354\\
O$_2$&21.4&6.2&3.7&32.9&9.6&5.6\\
\hline
\end{tabular}
\label{penetration}
\end{table}

Fayolle et al. \cite{Fayolle2} reported the photodesorption of N$_2$ and O$_2$ ice as a function of photon wavelength, see Figs. 1 and 3 of that work. It can be 
observed that the photodesorption in the 10.8-8.4 eV spectral range is very low compared with the photodesorption in the 12.0-13.8 eV range for N$_2$. This is mainly due to the 
low absorption cross section that we report in that region, see Fig.~\ref{N2}. For O$_2$, Fig.~\ref{O2} resembles the photodesorption profile of O$_2$ for different photon 
wavelengths in Fayolle et al. \cite{Fayolle2}. 

Using the expresion developed in Paper I,
\begin{equation}
R^{\rm abs}_{\rm ph-des} = \frac{I_0}{I_{abs}} \; R^{\rm inc}_{\rm ph-des},
\label{RR}
\end{equation}
where $I_0$ and $I_{abs}$ are the incident and the absorbed photon intensities at a certain wavelength, we can estimate the photodesorption 
rate per absorbed photon, $R^{\rm abs}_{\rm ph-des}$, which differs significantly from the photodesorption rate per incident photon, $R^{\rm inc}_{\rm ph-des}$, see Table \ref{Rabs}.

\begin{table}[ht!]
\centering
\caption{VUV-absorption cross sections of O$_2$ ice for different photon energies. R$^{\rm inc}_{\rm ph-des}$ values were adapted from Fig. 3 of Fayolle et al. (2013) for 
the starting, maximum, and minimum photon energies (7.6, 9.4, and 10.5 eV). The 8.6 eV value was added because it also coincides with the average photon 
energy of our MDHL. The values in the table correspond to 30 ML of O$_2$ ice (i.e., a column density of 30 $\times$ 10$^{15}$ molecules cm$[^{-2}$), see Fayolle et al. (2013).}
\tiny
\begin{tabular}{cccc}
%\aline
\hline
Irrad. energy&$\sigma$&R$^{\rm inc}_{\rm ph-des}$& R$^{\rm abs}_{\rm ph-des}$\\
eV&l cm$^{2}$& molec./photon$_{inc}$&  molec./photon$_{abs}$\\
\hline
%&&&\\
10.5& 7.1 $\times$ 10$^{-19}$&$\sim$ 1   $\times$ 10$^{-3}$& 0.05 \\
9.4 & 6.3 $\times$ 10$^{-18}$&$\sim$ 7   $\times$ 10$^{-3}$& 0.04 \\
8.6 & 8.0 $\times$ 10$^{-18}$&$\sim$ 4.5 $\times$ 10$^{-3}$& 0.02 \\
7.6 & 3.5 $\times$ 10$^{-18}$&$\sim$ 1.5 $\times$ 10$^{-3}$& 0.015\\
\hline
\end{tabular}
\label{Rabs}
\end{table}

For N$_2$ ice, Fayolle et al. \cite{Fayolle2} reported an upper limit of $R^{\rm inc}_{\rm ph-des}$ $\leq$ 4 $\times$ 10$^{-3}$ molecules per incident photon in the spectral range below 
12.4 eV for $N$ = 30 ML. The average absorption cross section for N$_2$ ice that we measured in that range is $\sigma$ = 7 $\times$ 10$^{-21}$ cm$^{2}$. 
The resulting photodesorption rate is quite high, $R^{\rm abs}_{\rm ph-des}$ $\leq$ 19 molecules per absorbed photon, meaning that a very small fraction of the incident photons 
are absorbed in the ice, but each absorbed photon led to the photodesorption of about 19 molecules on average (this in fact is a maximum value because Fayolle et al. \cite{Fayolle2} 
measured photodesorption rates that \emph{do not exceed} 4 $\times$ 10$^{-3}$ molecules per incident photon). 

A more direct comparison between N$_2$ and O$_2$ ice photodesorption rates could be made if the number of monolayers closer to the ice surface that truly contribute to the photodesorption 
were known. This value has not been estimated for other ices diferent from CO (about 5 ML, Mu\~noz Caro et al. 2010; Fayolle et al. 2011; and Chen et al. 2013). The above 
values of $R^{\rm inc}_{\rm ph-des}$ and $R^{\rm abs}_{\rm ph-des}$ correspond to the total ice column density of 30 ML in the experiment of Fayolle et al. (2013). With this 
uncertainty still remaining, we can conclude that when the VUV-absorption cross section of each specific ice composition is taken into account, it is possible to estimate the efficiency 
of the photodesorption 
per absorbed photon; in the case of N$_2$, for VUV photon energies that do not lead to direct dissociation of the molecules in the ice, these values are higher than unity. 
This observation and the fact that photons absorbed in ice monolayers deeper than the top monolayers (up to five for CO) can lead to a 
photodesorption event, indicate that the excess photon energy is transmitted to neighboring molecules in the ice within a certain range (Rakhovskaia et al. 1995; this range may 
correspond to about 5 ML in the case of CO ice, e.g. Mu\~noz Caro et al. 2010). In the case of O$_2$ the VUV photons have enough energy to dissociated the molecule, which makes 
it harder to measure the photodesorption rate because the dissociation dominates between the two processes.

\section{Conclusions}
\label{Conclu}

This work adds to Paper I to complete our set of nine molecular ice components (CO, H$_2$O, CH$_3$OH, NH$_3$, H$_2$S, CH$_{4}$, CO$_{2}$, N$_{2}$, and O$_{2}$) selected to perform 
VUV-spectroscopy in the 120-160 nm spectral range. Some key aspects of this work are summarized below.

\begin{itemize}
\item For the first time, to our knowledge, the VUV-absorption cross sections of CH$_4$, N$_2$, and O$_{2}$ were measured for the solid phase, with average 
VUV-absorption cross sections of 5.7 +0.5/-1.1 $\times$ 10$^{-18}$ cm$^{2}$, 7.0 +0.6/-1.4 $\times$ 10$^{-21}$ cm$^{2}$, and 4.8 +0.4/-1.0 $\times$ 10$^{-18}$ cm$^{2}$. 
The total integrated VUV-absorption cross sections are 2.0 +0.1/-0.4 $\times$ 10$^{-16}$ cm$^{2}$ nm,  2.3 +0.2/-0.5 $\times$ 10$^{-19}$ cm$^{2}$ nm, 
and 2.4 +0.2/-0.5 $\times$ 10$^{-15}$ cm$^{2}$ nm. Our estimated value of the average VUV-absorption cross section of CO$_{2}$ ice, 2.6 +0.2/-0.3 $\times$ 10$^{-17}$ cm$^{2}$, 
is comparable with that reported by Mason et al. \cite{Mason}, which was measured using a synchrotron as the monochromatic VUV-emission source.
\end{itemize}
\begin{itemize}
\item The ice samples made of N$_{2}$ or CO$_2$ display discrete VUV-absorption bands, the latter were not well resolved, while samples containing O$_{2}$ or CH$_{4}$ 
present a continuum VUV-absorption band, see Fig.~\ref{todas}, where N$_2$ presents by far the lowest VUV-absorption cross section. 
\end{itemize}
\begin{itemize}
\item The ice sample is inevitably UV-irradiated for a few minutes during the spectral acquisition in our experimental configuration, 
which can lead to photoproduct formation. This effect is clearly observed in the case of CO$_{2}$ because the CO photoproduced has a large 
VUV-absorption cross section compared with CO$_{2}$ in the solid phase. In this experiment, the CO bands are 
shifted in position and their profiles changed with respect to the pure CO ice spectrum reported in Paper I. These mixture effects were not significantly studied and have clear 
implications for the absorption of multicomponent ice mantles in space. 
\end{itemize}
\begin{itemize}
\item The VUV-absorption cross sections of the two homonuclear molecules studied, N$_{2}$ and O$_{2}$, differ by two orders of 
magnitude. This affects their photodesorption rates and the formation of photoproducts in the ice matrix.
\end{itemize}

\begin{acknowledgements}
This research was financed by the Spanish MICINN under projects AYA2011-29375 and CONSOLIDER grant CSD2009-00038. This work was partially supported by NSC 
grants NSC99-2112-M-008-011-MY3 and NSC99-2923-M-008-011-MY3, and the NSF Planetary Astronomy Program under Grant AST-1108898.
\end{acknowledgements}


\begin{thebibliography}{99}

\bibitem[(2003)]{Belloche} Belloche, A., \& André, P. 2003, ApJ, 593, 906
\bibitem[(2002)]{Bergin} Bergin, E. A., Langer, W. D., \& Goldsmith, P. F. 2002, ApJ Lett., 570, 101
\bibitem[(1978)]{Boursey} Boursey, E., Chandrasekharan, V., Gürtler, P., et al. 1978, Phys. Rev. Lett., 41, 1516
\bibitem[(1965)]{Brith} Brith, M. \& Schnepp, O., 1965, Mol. Phys., 9, 473.
\bibitem[(2010)]{Asper1} Chen, Y.-J., Chu, C.-C, Lin, Y.-C, et al., 2010, Adv. Geosci. 25, 259
\bibitem[(2013)]{Asper2} Chen, Y.-J., Chuang, K.-Y., \& Mu\~noz Caro, G. M. 2013, ApJ, in press.
\bibitem[(2006)]{Cheng} Cheng, B.-M., Lu, H.-C., Chen, H.-K., et al., 2006, ApJ, 647, 1535
\bibitem[(2011)]{Cheng2} Cheng, B.-M., Chen, H.-F., Lu, H.-C., et al., 2011, ApJ S.S., 196, 3
\bibitem[(1995)]{Chiar} Chiar, J. E., Adamson, A. J., Kerr, T. H., \& Whittet, D. C. B. 1995, ApJ, 455, 234
\bibitem[(2004)]{Collings} Collings, M. P., Anderson, M. A., Chen, R., et al., 2004, Mon. Not. R. Astron. Soc., 354, 1133
\bibitem[(2013)]{Cruz} Cruz-Diaz G. A., Mu\~noz Caro G. M., Chen Y.-J., 2013a, A\&A, accepted.
\bibitem[(1970)]{Dalgarno} Dalgarno, A. \& Stephens, T. L., 1970, ApJ, 160, L107
\bibitem[(1970)]{Darwent} Darwent, B. de B., 1970, National standard reference data system, 31, 70-602101
\bibitem[(2007)]{Dawes} Dawes, A., Mukerji, R.J., Davis, M. P., Holtom, P. D., \& Webb, S.M., 2007, J. Chem. Phys. 126, 244711
\bibitem[(1998)]{Ehrenfreund} Ehrenfreund, P. \& van Dishoeck, E. F., 1998, Adv. Space Res., 21, 15
\bibitem[(1999)]{Feng} Feng, R., Cooper, G., \& Brion, C. E., 1999, Chem. Phys., 244, 127
\bibitem[(2006)]{Fuchs} Fuchs, G. W., Acharyya, K., Bisschop, S. E., et al., 2006, Faraday Discussions, 133, 331
\bibitem[(2009)]{Fulvio} Fulvio, D., Sivaraman, B., Baratta, G. A., Palumbo, M. E., \& Mason, N. J., 2009, Spectrochimica Acta Part A, 72, 1007
\bibitem[(1975)]{Gallo} Gallo, A. R. \& Innes, K. K., 1975, J. Mol. Spectrosc., 54, 472
\bibitem[(1996)]{Gerakines} Gerakines, P. A., Schutte, W. A. \& Ehrenfreund, P., 1996, A\&A, 312, 289
\bibitem[(2011)]{Goldsmith} Goldsmith, P. F., Liseau, R., Bell, T. A., et al., 2011, ApJ, 737, 96
\bibitem[(1989)]{Gredel} Gredel, R., Lepp, S., \& Dalgarno, A., 1989, ApJ, 347, 289
\bibitem[(2011)]{Fayolle1} Fayolle, E. C., Bertin, M., Romanzin, C., Michaut, X., et al. 2011, ApJ Letters, 739, L36
\bibitem[(2013)]{Fayolle2} Fayolle, E. C., Bertin, M., Romanzin, C., Poderoso, H. A. M., et al. 2013, A\&A, 556, A122
\bibitem[(2011)]{Hincelin} Hincelin, U., Wakelam, V., Hersant, F., et al., 2011, A\&A, 530, A61
\bibitem[(1968)]{Hudson} Hudson, R. D. \& Carter, V. L., 1968, J. Opt. Soc. Am., 58, 227
\bibitem[(2002)]{Hudson2} Hudson, R. L. \& Moore, M. H., 2002, ApJ, 568, 1095
\bibitem[(1954)]{Inn} Inn, E. C. Y., 1954, Spectrochimica Acta, 7, 65
\bibitem[(2007)]{Jamieson} Jamieson, C. S. \& Kaiser, R. I., 2007, Chem. Phys. Letters, 440, 98
\bibitem[(2011)]{Antonio} Jim\'enez-Escobar, A. \& Mu\~noz Caro, G.M., 2011, A\&A. 536, 11
\bibitem[(2007)]{Kuo} Kuo, Y.-P., Lu, H.-C., Wu, Y.-J., Cheng, B.-M. \& Ogilvie, J. F., 2007, Chem. Phys. Lett., 447, 168
\bibitem[(2001)]{Lee} Lee, A. Y. T., Yung, Y. L., Cheng, B.-M., et al., 2001, ApJ Lett., 551, L93
\bibitem[(2012)]{Liseau} Liseau, R., Goldsmith, P. F., Larsson, B., Pagani, L., \& Bergman, P., 2012., A\&A, 541, A73
\bibitem[(2005)]{Lu} Lu, H.-C., Chen, H.-K., Cheng, B.-M., Kuo, Y.-P. \& Ogilvie, J. F., 2005, J. Phys. B: At. Mol. Opt. Phys., 38, 3693
\bibitem[(2008)]{Lu2} Lu, H.-C., Chen, H.-K., Cheng, B.-M. \& Ogilvie, J. F., 2008, Spectrochimica Acta Part A: Molecular and biomolecular spectroscopy, 71, 1485
\bibitem[(2010)]{Lu3} Lu, H. -C., Chen, H. -K., Chen, H. -F., Cheng, B. -M. \& Ogilvie, J. F., 2010, A\&A, 520, A19
\bibitem[(2010)]{Kalvans} Kalvans, J. \& Shmeld, I., 2010, A\&A, 521, A37
\bibitem[(2006)]{Mason} Mason, N. J., Dawes, A., Holton, P. D., et al., 2006, Faraday Discussions, 133, 311
\bibitem[(1976)]{Merrill} Merrill, K. M., Russell, R. W. \& Soifer, B. T., 1976, ApJ, 251, 533
\bibitem[(1998)]{Meyer} Meyer, D. M., Jura, M., \& Cardelli, J. A. 1998, ApJ, 493, 222
\bibitem[(1974)]{Monahan} Monahan, K. M. \& Walker, W. C., 1974, J. Chem. Phys., 61, 3886
\bibitem[(2005)]{Mota} Mota, R., Parafita, R., Giuliani, A., et al., 2005, Chem. Phys. Lett., 416, 152
\bibitem[(2011)]{Mumma} Mumma, M. J. \& Charnley, S. B. 2011, Annu. Rev. Astro. Astrophys., 49, 471
\bibitem[(2010)]{Caro1} Mu\~noz Caro, G. M., Jim\'enez-Escobar, A., Mart\'\i{}n-Gago, J. \'A., et al, 2010, A\&A, 522, A108
\bibitem[(1985)]{Nee} Nee, J. B., Suto, M., \& Lee, L. C., 1985, Chemical Physics, 98, 147.
\bibitem[(2007)]{Oberg1} \"Oberg K.I., Fuchs G.W., Awad Z. et al., 2007, ApJ, 662, L23
\bibitem[(2009)]{Oberg2} \"Oberg K.I., van Dishoeck E.F., \& Linnartz H., 2009, A\&A, 496, 281
\bibitem[(2011)]{Oberg} \"Oberg, K. I., Boogert, A. C. A., Pontoppidan, K. M., et al., 2011, The Molecular Universe, Proceedings IAU Symposium No 280
\bibitem[(1978)]{Okabe} Okabe, H., 1978, Photochemistry of small molecules, John Wiley \& Sons, New York
\bibitem[(2004)]{Pontoppidan2} Pontoppidan, K. M., van Dishoeck, E. F., \& Dartois, E. 2004, A\&A, 426, 925
\bibitem[(2008)]{Pontoppidan} Pontoppidan, K. M., Blake, G. A., van Dishoeck, E. F., et al., 2008, ApJ, 684, 1323
\bibitem[(1938)]{Price} Price, W. C. \& Simpson, D. M., 1938, Proc. Roy. Soc. (Lond.), A165, 272
\bibitem[(1995)]{Rakhovskaia} Rakhovskaia O., Wiethoff P., \& Feulner P., 1995, NIM B, 101, 169
\bibitem[(2000)]{Vacuum} Samson, J. A. R. \& Ederer, D. L., 2000, Vacuum Ultaviolet Spectroscopy, Elsevier Inc
\bibitem[(1988)]{Sandford2} Sandford, S. A., Allamandola, L. J., Tielens, A. G. G. M., \& Valero, G. J., 1988, ApJ, 329, 498
\bibitem[(1997)]{Sandford} Sandford, S. A., Allamandola, L. J., \& Bernstein, M. P., 1997, From stardust to planetesimals ASP conference series, 122, 201
\bibitem[(2008)]{Satorre} Satorre, M. \'A., Domingo, M., Mill\'an, C., et al., 2008, Planetary and Space Science, 56, 1748
\bibitem[(2002)]{Smith} Smith, P. L., Rufus, L., Yoshino, K., \& Parkinson, W. H., 2002, NASA Laboratory Astrophysics Workshop, NASA/CP-2002-21186, 158
\bibitem[(1979)]{Soifer} Soifer, B. T., Puetter, R. C., Russell, R. W., et al., 1979, ApJ, 232, L53
\bibitem[(1989)]{Sternberg} Sternberg, A., 1989, ApJ, 347, 863.
\bibitem[(1984)]{Tielens} Tielens, A. G. G. M., Allamandola, L. J., Bregman, J., et al., 1984, ApJ, 287, 697
\bibitem[(1991)]{Tielens2} Tielens, A. G. G. M., Tokunaga, A. T., Geballe, T. R., \& Baas, F. 1991, ApJ, 691, 1459
\bibitem[(1992)]{vanDishoeck} van Dishoeck, E. F., Phillips, T. G., Keene, J., \& Blake, G. A., 1992, ApJ Lett., 441, 222
\bibitem[(1996)]{Whittet} Whittet, D. C. B., Schutte, W. A., Tielens, A. G. G. M., et al., 1996, A\&A, 315, L357
\bibitem[(2007)]{Wu2} Wu, Y.-J., Lu, H.-C., Chen, H.-K., \& Cheng, B.-M., 2007, J. Chem. Phys., 127, 154311
\bibitem[(2010)]{Wu3} Wu, Y.-J., Lin, M.-Y., Chou S.-L., et al., 2010, ApJ, 721, 856
\bibitem[(2012)]{Wu} Wu, Y.-J., Wu, C. Y. R., Chou S.-L., et al., 2012, ApJ, 746, 175
\bibitem[(1996)]{Yoshino} Yoshino, K., Esmond, J. R., Sun, Y., et al., 1996, J. Quant. Spectrosc. Radiat. Transfer, 55, 53
\bibitem[(1964)]{Yamada} Yamada, H. \& Person, W. B. 1964, J. Chem. Phys., 41, 2478
\end{thebibliography}
\end{document}